\documentclass[prl,aps,twocolumn,superscriptaddress]{revtex4}
\usepackage{graphicx}
\usepackage{dcolumn}
\usepackage{bm}
\usepackage[dvips]{color}
\newcommand{\br}{{\bf r}}

\begin{document}

\title{Supersolid Order from Disorder: Hard-Core Bosons on the 
Triangular Lattice}
\author{R.~G.~Melko}
\affiliation{Department of Physics, University of California, Santa Barbara, 
California 93106}
\author{A.~Paramekanti}
\affiliation{Department of Physics, University of California, Berkeley, 
California 94720} 
\author{A.~A.~Burkov}
\affiliation{Department of Physics, University of California, Santa Barbara, 
California 93106}
\author{A.~Vishwanath}
\affiliation{Department of Physics, University of California, Berkeley, 
California 94720}  
\author{D.~N.~Sheng}
\affiliation{Department of Physics and Astronomy, California State University,
Northridge, California 91330}
\author{L.~Balents}
\affiliation{Department of Physics, University of California, Santa Barbara, 
California 93106}
\date{\today}

\begin{abstract}
  We study the interplay of Mott localization, geometric frustration,
  and superfluidity for hard-core bosons with nearest-neighbor
  repulsion on the triangular lattice.  For this model at
  half-filling, we demonstrate that superfluidity survives for
  arbitrarily large repulsion, and that diagonal solid order emerges
  in the strongly correlated regime from an order-by-disorder
  mechanism.  This is thus an unusual example of a stable supersolid
  phase of hard-core lattice bosons at a commensurate filling.
\end{abstract}

\maketitle

{\it Introduction}: Recent experiments on $^4$He under pressure
suggest \cite{ssHe} that a supersolid phase, in which long-range
diagonal (crystalline) order and long-range off-diagonal (superfluid)
order coexist, may arise at temperatures $T \lesssim 100$ mK.  Such a
phase had been theoretically envisioned long ago
\cite{Andreev69}, based on the proposition that a
nonvanishing density of zero point defects in the solid (interstitials
or vacancies) could Bose condense and form a superfluid ``on top'' of
the existing crystalline background. However, early estimates
\cite{Leggett70} of the superfluid fraction in clean $^4$He solid were
very small ($\sim 10^{-4}$), and recent numerics with realistic
interatomic potentials find no evidence for zero point defects or
appreciable interparticle exchanges \cite{Ceperley04} in the crystal.

Since the microscopic conditions under which clean $^4$He might exist
in a supersolid phase are still unclear, it is useful to focus on
simple lattice models of bosons in order to understand the different
mechanisms by which a supersolid might emerge. Such ``lattice
supersolids'' are superfluids which also break the lattice translation
symmetries \cite{Fisher73}.  Numerical studies of interacting bosons
on the square lattice find that stable supersolid phases form upon
doping away from a half-filled checkerboard or striped solid
\cite{Otterlo94,Batrouni00,Pryadko05}. These appear to be examples of
the ``defect condensation'' mechanism, where doped bosons (holes) act
as interstitial (vacancy) sites in the crystal.

In this letter we explore a different route to supersolidity based on
the competition between Mott localization physics and geometric
frustration. To illustrate this, we focus on a simple model of
hard-core bosons at half-filling on the triangular lattice, interacting
via a nearest-neighbor repulsive term,
\begin{equation}
\label{eq:1}
H = - t \sum_{\langle ij \rangle}\left(b^{\dag}_i b^{\vphantom\dag}_j + b_j^{\dag} b^{\vphantom\dag}_i \right)
+ \sum_{\langle ij \rangle} V (n_i-\frac{1}{2}) (n_j-\frac{1}{2}).
\end{equation}
For this model, we show that a supersolid phase emerges for
$V/t \gtrsim 10$.
The supersolid phase arises from an order-by-disorder effect in a
strongly correlated superfluid. This route to supersolidity is thus
complementary to the ``defect condensation'' mechanism which starts
from a perfect crystal and considers a small defect-density induced
supersolid.   Our results are based on a combination of analytical
arguments, projected boson wavefunctions, quantum Monte Carlo and
exact diagonalization studies.  We thus go well beyond the earlier
spin-wave work \cite{Murthy97} on this model. 
It would be interesting
to look for signatures of this supersolid in experiments, e.g on
interacting cold bosonic atoms in an optical lattice \cite{DDLprl}, 
where both a zero momentum condensate peak and Bragg peaks at the ordering
wavevectors would be expected in the usual time-of-flight momentum
spectroscopy.

\begin{figure}
\includegraphics[width=9.8cm]{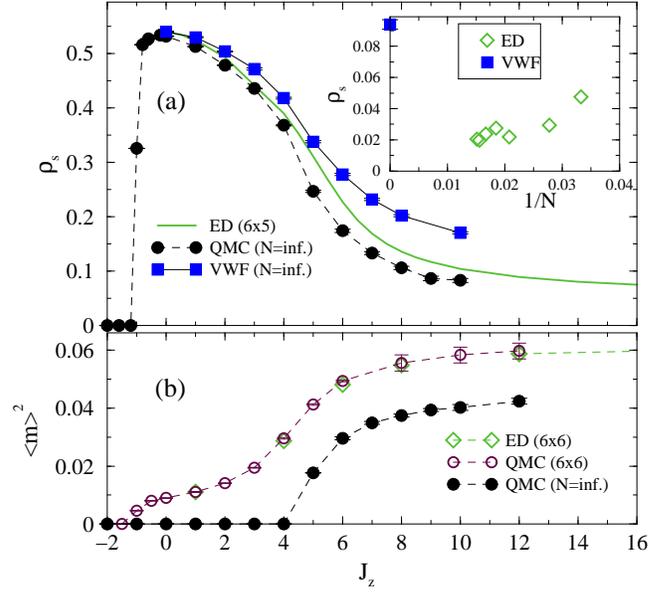}
\vskip -0.2cm
\caption{(color online). 
(a) Ground state superfluid density, 
calculated with exact diagonalization, quantum Monte Carlo, and variational 
wavefunction, with $J_{\perp}=1$.  
The inset is $\rho_s$ in the limit $J_z \rightarrow \infty$, 
calculated by ED (for several system
sizes $N$), and VWF (in the thermodynamic limit). 
(b) The diagonal order parameter squared.
Examples of the QMC extrapolations to $N\rightarrow \infty$ 
are illustrated in Fig.~\ref{FSSfig}.  
}
\label{MCrhos}
\vskip -0.5cm
\end{figure}

Alternatively, the Hamiltonian we study can be viewed, via the
standard mapping from hardcore bosons to S=1/2 spins, as an XXZ model:
\begin{equation}
H = \sum_{\langle ij \rangle} \left[- J_{\perp} 
\left( S^x_i S^x_j+S^y_i S^y_j \right)
+ J_z S^z_i S^z_j\right],
\label{eq:2}
\end{equation}
with $J_z=V$ and $J_\perp=2 t$.  A spin Hamiltonian such as Eq.~(\ref{eq:2})
may be realized in a Mott insulating phase of cold bosons on an
optical lattice, if the bosons possess two internal states which can
play the role of the spin \cite{DDLprl}. Since the two-body exchange
interaction for spinful bosons is naturally ferromagnetic, $J_\perp
>0$ in Eq.~(\ref{eq:2}).  In spin language, supersolid ordering corresponds to
the spins having their $xy$-components aligned ferromagnetically
(superfluid) with the $z$-component also ordered at a nonzero
wavevector (solid). In our analysis, we will work interchangeably in
terms of bosons or spin variables.


{\it Spin wave theory and Landau-Ginzburg-Wilson approach for
  $J_\perp/J_z \gg 1$}:~ For large $J_\perp/J_z$, the spins align
in-plane to form an XY-ferromagnetic phase. Ignoring small quantum
fluctuations, the ground state has all spins uniformly polarized along
the (say) $x$-direction, giving a non-zero $\langle S^x \rangle$. This
corresponds to a Bose condensate of the bosons, with off-diagonal long
range order and a nonzero superfluid stiffness.  Doing a spin wave
expansion about this ordered state, we find that the gap to
excitations at wavevectors $\pm {\bf k}_0= \pm (4\pi/3,0)$
(Brillouin zone corners) closes at $J_z/J_\perp=2$. In the boson language,
the ``roton minimum'' of the superfluid hits zero energy.  The uniform
superfluid develops an instability toward $3$-sublattice ordering,
$S_i^z \sim {\rm Re}( \psi e^{i{\bf k}_0\cdot{\bf r}_i})$, with the
order parameter $\psi$ describing the critical modes.  The
Landau-Ginzburg-Wilson (LGW) expansion of the effective action in $\psi$
and the {\sl uniform} Ising magnetization $M$ has the form:
\begin{eqnarray}
\label{LGW}
\mathcal{S}&=&\int d^2 x \int_0^{\beta} d \tau \left[ |\partial_{\tau} \psi|^2 
+ c^2 |{\boldsymbol \nabla} \psi|^2 + r |\psi|^2 + u |\psi|^4 \right.
\nonumber \\ 
&+&\left.v |\psi|^6 + w {\rm Re}\,(\psi^6) + M^2/(2\chi) - \lambda M
  \,{\rm Re}\, (\psi^3)\right],
\end{eqnarray}
where the constant $\chi$ represents the boson compressibility, and
$\lambda$ is generally non-zero from symmetry considerations.
Based as it is upon the spin-wave expansion, this LGW action describes
the onset of non-zero $\psi$ {\sl within} the superfluid, i.e. the
ordered state is a {\sl supersolid} (as opposed to an insulating solid
with a broken lattice symmetry).   Exhaustive evidence supporting this
interpretation will be provided below.  

The ``massive'' $M$ field can be integrated out to
obtain an effective action for $\psi$ alone, with the renormalized
parameters $\tilde{v}=v-\chi\lambda^2/2$, $\tilde{w}=w-\chi\lambda^2/2$.
Two possible ordered states thereby occur, determined by the sign of
$\tilde{w}$.  For $\tilde{w}<0$, $\psi^3$ is purely real, and from
Eq.~(\ref{LGW}), a non-zero spontaneous uniform magnetization $M$ is
predicted.  This {\sl ferrimagnetic} state has a three-sublattice
structure with $\langle S_i^z\rangle =(m,m,-m')$ ($M=(2m-m')/3\neq 0$),
and requires phase separation in the canonical boson number
ensemble.  For $\tilde{w}>0$, $\psi^3$ is purely
imaginary \cite{Berker84}, and $M=0$.  The corresponding
three-sublattice pattern has $\langle S_i^z\rangle =(m,-m,0)$, so we
refer to this as an {\sl antiferromagnetic} state (no phase separation
is implied).  
Note that the instability to three-sublattice ordering was also found
by Murthy {\it et al.} \cite{Murthy97} who carried out a study of the
fluctuations around such possible ordered states within spin wave
theory, and suggested a supersolid phase for large-S.  Here we provide
a more thorough and rigorous analysis, directly for the $S=1/2$ case
of most interest.  Before proceeding, we note that Eq.~(\ref{LGW}) is an
appropriate starting point for investigating the quantum critical
behavior at the zero temperature superfluid-supersolid transition,
which is expected to be of the three-dimensional XY universality
class.

{\it Variational wavefunction}:~ A
simple variational wavefunction argument indicates that superfluidity
survives in our model for arbitrarily small $J_{\perp}/J_z$.  A good
variational approximation to the ground state for {\sl large}
$J_\perp/J_z$ is the hard-core projection of the non-interacting
Bose-condensate wavefunction,
\begin{eqnarray}
\Psi_0(\br_1,...,\br_{N_b}) = \left\{ \begin{array}{ll}
1 & \mbox{all distinct $\br_i$} \label{VWF1} \\
0 & \mbox{otherwise}
\end{array}
\right. ,
\end{eqnarray}
where $N_b=N/2$, and $N$ is the number of sites.  To account for
inter-site correlations with increasing $J_z$, we modify the
variational wavefunction in Eq.~(\ref{VWF1}) through a Jastrow factor
\begin{equation}
\Psi(\{\br\},g)= \left[ \prod_{\langle i,j\rangle} (1-g)^{(n_i-\frac{1}{2})
(n_j-\frac{1}{2})} \right] \Psi_0(\{\br\}),
\label{varwfn}
\end{equation}
where $0 \leq g < 1$ is a variational parameter, and $\{\br\}
\equiv (\br_1\ldots\br_{N_b})$. 
Using a Monte Carlo method, we have evaluated correlation functions in 
$\Psi$ with the optimal $g \equiv g^*$ at each value of $J_z/J_\perp$.

For $J_z=0$, we find $g^*\approx 0.2$, and the wavefunction in 
Eq.~(\ref{varwfn}) has off-diagonal long range order, with
$\langle b_i \rangle \approx 0.48$. Imposing a phase gradient $\phi$
in the wavefunction $\Psi$ through a twist in the boundary condition,
the superfluid stiffness is defined as
\begin{equation}
\rho_s = \frac{\partial^2 E}{\partial \phi^2} 
= \lim_{\phi \rightarrow 0} \frac{\Delta E}{N} \frac{2}{\phi^2},
\label{RHOs}
\end{equation}
where $\Delta E$ is the resulting change in energy.  Using this definition,
we estimate $\rho_s(J_z=0) \approx 0.54 J_\perp$. With
increasing $J_z$, we find $g^*$ increases and the superfluid density
decreases, as illustrated in Fig.~\ref{MCrhos}a.
  
Let us go to the most interesting limit of $J_\perp/J_z \to 0$.  For
$J_\perp=0$, the XXZ model in Eq.~(\ref{eq:2}) reduces to the classical Ising
antiferromagnet on a triangular lattice which is
well-known to have a macroscopic degeneracy of ground states $\approx
1.381^{N}$. These are configurations with exactly
one frustrated bond per triangle \cite{ClassIsing}.  For
$J_z\to \infty$, we find the optimal $g^* \to 1$, and the
wavefunction $\Psi(\{\br\},0)$ is then an equal weight linear
superposition of all Ising ground state configurations with $S^z_{\rm
tot}=0$.  In this case, we find the off-diagonal long
range order $\langle b_i \rangle \approx 0.13$, and superfluid
stiffness $\rho_s \approx 0.09 J_\perp$ (see Fig.~\ref{MCrhos}a).  This
persistence of off-diagonal long range order on the triangular lattice
for $J_{\perp}/J_z \rightarrow 0$ is one of the most important points
of this paper and is crucial for the discussion below. 

Remarkably, despite its simplicity, the variational wavefunction with
$g \to 1$ can also be shown (analytically 
and numerically) to have power-law $1/\sqrt{r}$ correlations of $S_i^z$ at
the wavevectors $\pm{\bf k}_0$, in addition to off-diagonal order,
reminiscent of the long-range Ising correlations expected from the LGW
theory.  Since it somewhat underestimates the ``solid'' order, we
expect that the variational approximation slightly overestimates
superfluidity, as also observed numerically.


{\it Strong coupling analysis for $J_\perp/J_z\! \ll\! 1$}:~ We
now turn to directly consider the strong-coupling limit.  A simple
qualitative picture (which we will reinforce with a rigorous analysis
a little later) can be obtained by mapping the XXZ model onto a
transverse field Ising model:
\begin{equation} 
H = J_z \sum_{\langle ij \rangle} S^z_i S^z_j - h_{\rm eff} \sum_i S^x_i, 
\label{eq:9} 
\end{equation} 
with $h_{\rm eff} = J_\perp \langle S^x_i \rangle$.  We have assumed
here that, as suggested by the variational wavefunction analysis,
superfluidity survives in the limit of small $J_{\perp}/J_z$.  This
transverse field Ising model has been studied earlier
\cite{Berker84,Moessner1} in various contexts, using Ginzburg-Landau
theory and Monte Carlo numerics.  It is well established that this model
has diagonal long-range order, in particular of antiferromagnetic type,
for small $h_{\rm eff}/J_z$.  This analogy suggests that the supersolid
behavior of the XXZ model should be viewed, similarly to
Eq.~(\ref{eq:9}), as arising microscopically from an order-by-disorder
mechanism~\cite{kagome}.

One might worry that the above analysis was based on a mean-field
treatment of the superfluid component of the order, which may be
unreliable at small $J_{\perp}/J_z$.  We now turn to a more exact
argument, which strongly argues for a ground state with broken spatial
symmetries.  In particular, we prove that for $J_{\perp}/J_z
\rightarrow 0$, the ground states are three-fold degenerate and
transform non-trivially under lattice rotations.  

In this limit ($J_z=\infty$), the XXZ model reduces to
\begin{equation}
H_\infty = -J_\perp \sum_{\langle ij\rangle} \hat{P}_{C}  (S_i^x
S_j^x+S_i^yS_j^y) \hat{P}_{C},\label{eq:CIAF}
\end{equation}
where $\hat{P}_{C}$ is a projection operator onto the classical Ising
antiferromagnetic ground state manifold.  
Consider an $L\times L$ system with periodic boundary conditions, 
i.e.~on the torus.  We define the quantities $E_a\geq 0$, $a=1,2,3$, as the
number of frustrated bonds along a straight closed path (loop) through $L$
sites along the $a^{th}$ principal axis of the triangular lattice.
These quantities are independent of translations of the loop.
Moreover, it is straightforward to show that they commute with
$H_\infty$ in Eq.~(\ref{eq:CIAF}) (and indeed any local Hamiltonian in
this manifold).  Because each triangle has one frustrated bond, one
also finds $E_1+E_2+E_3=L$.  Under a six-fold rotation of the system,
the $E_a$ cyclically permute.  Hence a unique ground state with
$E_1=E_2=E_3$ that is also rotationally invariant can be found only
for $L$ a multiple of three.  For $L\neq 0 ({\rm mod}\, 3)$, the ground
state thus forms at least a three-fold degenerate rotational
multiplet, as promised (it can be spanned by three states with
``angular momentum'' $\ell=0,1,2\, ({\rm mod}\,3)$).

More refined arguments 
indicate that this degeneracy
is split only by terms of $O(e^{-cL})$ (with $c$ independent of $L$)
for $0<J_\perp/J_z\ll 1$, and $L\rightarrow \infty$.  Hence a
non-trivial rotational multiplet remains the ground state with
exponential accuracy in the thermodynamic limit.  This behavior is
inconsistent with a simple uniform superfluid, which is expected to
have only states with excitation energies above the ground state
scaling as a power-law of $L$.  The most natural possibility
consistent with this behavior is broken spatial symmetry of the ground
state.  Other arguments can be presented against more ``exotic''
possibilities.
Having concluded that the system
must break lattice symmetries, and already argued from our variational
wavefunction that superfluidity persists for $J_z \to \infty$, we thus
come to the conclusion that the ground state at large $J_z$ is a
supersolid, as our numerical simulations below prove.

\begin{figure}
\includegraphics[width=7.4cm]{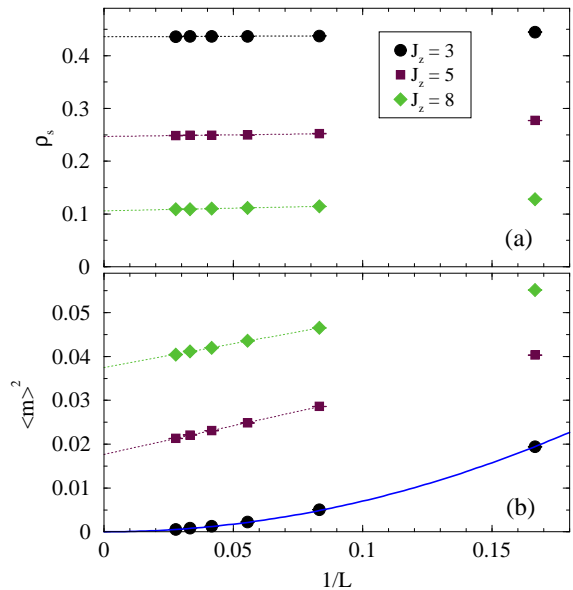}
\vskip -0.3cm
\caption{
(color online).
Finite-size scaling of QMC data for $\rho_s$ (a) and
$\langle m \rangle^2 = S({\bf k}_0)/L^2$ (b).
Here, $J_{\perp}=1$ and the inverse temperature is $\beta = 10$.
Error bars (illustrated) are much smaller than the symbol sizes.
Dotted lines show linear extrapolations for the five
largest system sizes.  The thick solid
(blue) line is a fit to $\langle m \rangle^2 \propto 1/L^2$,
expected asymptotically when there is no diagonal LRO.
}
\label{FSSfig}
\vskip -0.4cm
\end{figure}
{\it Exact numerical results:}~ To address the properties of the
spin-1/2 XXZ model in Eq.~(\ref{eq:2}) 
we have carried out exact diagonalization (ED) studies, as
well as Stochastic Series Expansion (SSE) quantum Monte Carlo (QMC)
simulations \cite{AWSsse} in the grand canonical framework.
Using straightforward modifications \cite{Serg} to the original SSE
directed-loop algorithm \cite{directed}, the QMC is able to explore all 
regions of the finite-temperature phase diagram for small $J_z/J_{\perp}$. 
However for $J_z/J_{\perp} \gtrsim 10$, the QMC is observed to experience a dynamical 
freezing due to the development of large energy barriers for local updates.
In Fig.~\ref{MCrhos}, we summarize our results for ED over the entire range of 
coupling strengths, and QMC in the weak to intermediate coupling regime 
(where we have been able to reliably equilibriate data).

The superfluid density is estimated by the 
stiffness, Eq.~(\ref{RHOs}), easily measured using 
winding numbers in the QMC simulations \cite{directed,pollock}. 
For a given $J_z/J_{\perp}$, $\rho_s$ 
is extrapolated to $L \rightarrow \infty$ using finite-size scaling 
on systems of size $N=L \times L$ (Fig.~\ref{FSSfig}a).
The results show that for small $J_z/J_{\perp}$ the system is 
clearly in a robust superfluid ground state with a large $\rho_s$.
The stiffness value decreases with increasing $J_z/J_{\perp}$, 
however it remains finite to the largest values of $J_z/J_{\perp}$ 
studied with the QMC.  The survival of $\rho_s$ at large $J_z/J_{\perp}$
is in stark contrast to the behavior at $J_z/J_{\perp} = -1$, 
where the phase transition to a fully-polarized ferromagnetic phase 
is evident by the vanishing stiffness.
Further, ED calculations (Fig.~\ref{MCrhos}a, inset) strongly 
suggest that $\rho_s$ survives to $J_z/J_{\perp} \rightarrow \infty$,
although system sizes are too small to allow a reliable
extrapolation to $N \rightarrow \infty$ for this data.
The small value of $\rho_s$ for $J_z \rightarrow \infty$
indicates that the supersolid is very close in 
parameter space to an insulating state.

In the above analytical arguments, the survival of $\rho_s$ at
large $J_z/J_{\perp}$ is accompanied by the development 
of simultaneous Ising order (the supersolid phase).
To address this prediction, we compute the static structure factor
\begin{equation}
S({\bf k}) = \frac{1}{N} \sum_{\langle ij \rangle} 
{\rm e}^{i\,{\bf k}\cdot({\bf r}_i - {\bf r}_j)} 
\langle S^z_i S^z_j \rangle.
\label{Szstr}
\end{equation}
This structure factor reveals the development of sharp 
Bragg peaks for moderate values of $J_z/J_{\perp}$.
The peaks occur at the corners of the hexagonal Brillouin
zone ($\pm {\bf k}_0= \pm (4\pi/3,0)$), in complete agreement 
with the above analytical discussion.
The presence of diagonal long range order (LRO) can be 
demonstrated rigorously through a finite-size scaling 
analysis of $S({\bf k}_0)$ (Fig.~\ref{FSSfig}b).  Specifically, in
a LRO state, the diagonal order parameter squared,
$\langle m \rangle^2 = S({\bf k}_0)/L^2$, should scale to a constant
as $L \rightarrow \infty$ (with a leading order-correction 
varying as $1/L$ \cite{Huse}).  
In contrast, in the absence of 
diagonal LRO, spin-spin correlations are short ranged, and
$\langle m \rangle^2$ should tend to zero as $1/L^2$ in
the asymptotic limit.  
Employing this analysis on the QMC data for all $J_z/J_{\perp}$
studied reveals the position of the zero-temperature phase 
transition from the superfluid to supersolid state to be between 
$J_z/J_{\perp}=4$ and $J_z/J_{\perp}=5$ (Fig.~\ref{MCrhos}).

Finally, we note that the grand canonical QMC simulations indicate no
significant deviations from half-filling in the supersolid phase
for system sizes up to $L=36$.  
In the ED, the minimum-energy ground state is also 
found to lie within the $\langle S^z \rangle = 0$ subspace.
These observations suggest the occurrence of the antiferromagnetic 
rather than the ferrimagnetic supersolid, however QMC simulations on 
much larger system sizes may be necessary in order to distinguish 
conclusively between the two.


{\it Conclusions:} We have presented a comprehensive study of the
phase diagram of interacting hard-core bosons at half-filling (zero
field XXZ model) on the triangular lattice. We have found that a
subtle interplay between repulsion and quantum fluctuations produces a
supersolid ground state in this system in the strongly correlated
regime.  This model provides an unusual example of a supersolid at a
commensurate density arising from an order-by-disorder mechanism
rather than defect condensation.  This distinct mechanism completely
avoids the instability of the {\sl defect-mediated} square lattice
supersolid to phase separation \cite{Pryadko05}. Here, the supersolid
has boson density equal to or extremely close to $1/2$, which is
strongly favored by the maximal frustration-induced entropy of the
corresponding classical Ising antiferromagnetic states.

\acknowledgments{
The authors would like to thank A.~W. Sandvik, D.~J.~Scalapino and 
D. Stamper-Kurn for insightful discussions.  
Supercomputer time was provided by NCSA under grant number DMR020029N.
Financial support was provided by National Science Foundation Grants
DMR02-11166 (RGM), DMR-9985255 (LB,AAB),  DMR-0307170 (DNS), 
the Packard Foundation (LB,AAB), ACS-PRF 41752-AC10 (DNS) and 
DOE LDRD DE-AC03-76SF00098 (AP and AV). 
During completion of this work we became aware of parallel numerical 
works \cite{damle}, which reach similar conclusions.


\begin{thebibliography}{99}
\bibitem{kagome} In contrast, generalizations of our analytical
arguments predict a simple {\sl uniform} superfluid on the kagome
lattice at all $J_z/J_\perp$.  This prediction has recently
been confirmed by QMC calculations \cite{Serg}.
\end{thebibliography}
\end{document}